\documentclass[prl,twocolumn,showpacs,superscriptaddress,showkeys]{revtex4}
\usepackage{amsmath,amssymb,amsthm,times,graphics,graphicx,bm}
\usepackage[colorlinks,linkcolor=red]{hyperref}

\newcommand{\be}{\begin{equation}}
\newcommand{\ee}{\end{equation}}
\newcommand{\ben}{\begin{eqnarray}}
\newcommand{\een}{\end{eqnarray}}
\newcommand{\bes}{\begin{subequations}}
\newcommand{\ees}{\end{subequations}}
\newcommand{\dg}{\dagger}

\def\ket#1{ | #1 \rangle}
\def\bra#1{{\langle #1 | }}
\def\tr{ {\rm{Tr }}}

\newcommand{\proj}[1]{\mbox{$|#1\rangle \!\langle #1 |$}}

\begin{document}
\title{Quantum discord between relatively accelerated observers}
\author{Animesh Datta}
\email{a.datta@imperial.ac.uk}
 \affiliation{Institute for Mathematical Sciences, 53 Prince's Gate, Imperial College, London, SW7 2PG, UK}
 \affiliation{QOLS, The Blackett Laboratory, Prince Consort Road, Imperial College, London, SW7 2BW, UK}

\date{\today}
\begin{abstract}
We calculate the quantum discord between two free modes of a
scalar field which start in a maximally entangled state and then
undergo a relative, constant acceleration. In a regime where there
is no distillable entanglement due to the Unruh effect, we show
that there is a finite amount of quantum discord, which is a
measure of purely quantum correlations in a state, over and above
quantum entanglement. Even in the limit of  infinite acceleration
of the observer detecting one of the modes, we provide evidence
for a non-zero amount of purely quantum correlations, which might
be exploited to gain non-trivial quantum advantages.
\end{abstract}



\maketitle


The theory of relativity and quantum theory, together with
information theory may be said to form the cornerstones of
theoretical physics~\cite{pt04}. The last of these two are being,
over the last decade, amalgamated into the field of quantum
information science that seeks to compute and process information
limited by the laws of quantum mechanics~\cite{nielsen00a}. The
enterprise of incorporating the principles of the theory of
relativity into quantum information is, in comparison, nascent.
Nonetheless, there have been several studies at the intersection
of relativity theory and quantum information science, particularly
in the study of Bell's
inequalities~\cite{c97,crw09,ks05,ms06,tu02}, quantum
entropy~\cite{pst02,c05}, quantum
entanglement~\cite{gba03,ga02,s04,sm05,jss07},
teleportation~\cite{am03}, and beyond~\cite{bd08}. There have also
been studies involving the entanglement in fermionic
fields~\cite{asmt06} and continuous-variable systems in
nonintertial frames~\cite{ase07}. These have shown that
entanglement between some degrees of freedom can be transferred to
others, and that the notion of entanglement is observer dependant.

        In addition to the investigations into fundamental nature
of quantum entanglement in a curved space-time, there have been
several proposals for detecting relativistic effects in laboratory
systems like cavity QED~\cite{skbfc03}, ion traps~\cite{adm05} and
atom dots in Bose-Einstein condensates~\cite{rcpr08}. These
effects of detecting acceleration radiation is a consequence of
the Unruh effect~\cite{chm08}. A result from quantum field theory,
it states that uniformly accelerated observers (that is, with
constant proper acceleration) in Minkowski space-time associate a
thermal bath to the vacuum state of the inertial observers. For
the inertial observer, the Minkowski coordinates $(T,Z)$ are
appropriate, while for a uniformly accelerating observer, Rindler
coordinates $(\tau,\xi)$ are more apt. Minkowski space-time is
invariant under the boosts, and this motivates the hyperbolic
coordinate transformations
    \ben
    T &=& \frac{1}{a}e^{a\xi}\sinh a\tau,\;\;Z=\frac{1}{a}e^{a\xi}\cosh
    a\tau,\;\;|Z|<T,\\
    T &=& -\frac{1}{a}e^{a\xi}\sinh a\tau,\;\;Z=\frac{1}{a}e^{a\xi}\cosh a\tau,\;\;|Z|>T.
    \een
These two transformations lead to two sets of Rindler coordinates,
called the right and left Rindler wedges respectively, which
together form a complete set of solutions of the Klein-Gordon
equation in Minkowski space-time.

The solutions of the Klein-Gordon equation in Minkowski space-time
are related to those in the Rindler wedges via a Bogoluibov
transformation~\cite{chm08}. These transform the vacuum of the
inertial observer into a two-mode squeezed state for the
accelerating observer, the two modes residing in the two Rindler
wedges. If we probe only one of the wedges, as we are constrained
to due to causality, the other mode is traced over, leaving us
with a mixed state of free bosons at a temperature proportional to
the acceleration. Additionally, if one starts with a pure
entangled state of two free modes of a scalar field shared between
two observers, Alice and Bob, and one of them, say Bob
accelerates, the result is a mixed state, whose entanglement, as
measured by the logarithmic negativity, is degraded from the point
of view of Rob (accelerating Bob)~\cite{sm05}, while there is no
change from the point of view of Alice.

Our endeavor in this paper will be to explore the above phenomenon
from the perspective of quantum
discord~\cite{ollivier01a,henderson01a,dattathesis}. It is a
measure of purely quantum correlations, and we show that although
the quantum discord suffers some degradation, there is a finite
amount of quantum discord between Alice and Rob at accelerations
at which the distillable entanglement has gone to zero. The use of
quantum discord is firstly motivated by the fact that noninertial
observers inevitably encounter mixed states, for which there is a
lack of universally accepted, easily computable measures of
entanglement. Quantum discord is ideally suited for application to
mixed states. Secondly, quantum discord is a measure of purely
quantum correlations, over and above entanglement, although for
pure states, they coincide. Finally, the quantum discord has been
presented as a possible resource for certain quantum
advantages~\cite{datta08a}, and the presence of nonzero amounts of
quantum discord as perceived by the nonintertial observer might
allow him to achieve nontrivial quantum advantage beyond a point
where the distillable entanglement touches zero. We will also show
that a `symmetrized' form of quantum discord, called the MID
(measurement induced disturbance) measure~\cite{Luo08a,dg09}, and
defined as the difference between the entropy of a quantum state,
and that obtained by measuring both the subsystems in their
reduced eigenbases, has a finite value at accelerations at which
the logarithmic negativity is zero. This is comparatively easier
to calculate than the quantum discord, and is an upper bound on
it. What both these measures however show, is, that starting with
an initially entangled state shared between Alice and Bob, there
will persist quantum correlations between them when Bob
accelerates, beyond accelerations at which the distillable
entanglement has fallen to zero.

For concreteness, we start with the maximally entangled state
between Alice and Bob, of two Minkowski modes $s$ and $k$
 \be
 \label{bell}
 \ket{\Psi}^{\mathcal{M}}=\frac{1}{\sqrt2}(\ket{0_s}^{\mathcal{M}}\ket{0_k}^{\mathcal{M}}+\ket{1_s}^{\mathcal{M}}\ket{1_k}^{\mathcal{M}}).
 \ee
When Bob accelerates with respect to Alice with a constant
acceleration, the Minkowski vacuum can be expressed as a two-mode
squeezed state of the Rindler vacuum~\cite{chm08}
 \be
 \ket{0_k}^{\mathcal{M}}=\frac{1}{\cosh r}\sum_{n=0}^{\infty}\tanh^n\!r\ket{n_k}_{1}\ket{n_k}_{2}
 \ee
with
 \be
 \tanh r = e^{-\pi|k|c/a}\equiv t,
 \ee
and $\ket{n_k}_1$ and $\ket{n_k}_2$ refer to the two modes,
corresponding to the left and right Rindler wedges. An excitation
in the Minkowski mode can be easily represented as
 \be
 \ket{1_k}^{\mathcal{M}}=\frac{1}{\cosh r}\sum_{n=0}^{\infty}\tanh^n\!r\sqrt{n+1}\ket{(n+1)_k}_{1}\ket{n_k}_{2}.
 \ee
As only one of the modes is accessible to Rob due to the causality
constraint, modes in one of the Rindler wedges (say mode 2) need
to be traced over. Using the above expressions, the maximally
entangled state in Eq. (\ref{bell}) is now transformed into
 \ben
 \label{rhoar}
 \rho_{AR} &=& \frac{1}{2\cosh^2r}\sum_{n=0}^{\infty}\tanh^n\!r\rho_n,  \\
 \mbox{where} && \nonumber \\
 \rho_n &=& \proj{0,n} + \frac{\sqrt{n+1}}{\cosh r}\ket{0,n}\bra{1,n+1} \nonumber\\
 &+&\!\!\!\frac{\sqrt{n+1}}{\cosh r}\ket{1,n+1}\bra{0,n} +
 \frac{n+1}{\cosh^2r}\proj{1,n+1}\nonumber
 \een
with $\ket{m,n}\equiv\ket{m_s}^{\mathcal{M}}\ket{n_k}_1.$ The
entanglement in the state Eq. (\ref{rhoar}) shared by Alice and
Rob has been calculated in Ref~\cite{sm05}. Our aim in this paper
will be to calculate the quantum discord in this state.


Quantum discord aims at capturing all quantum correlations in a
state, including entanglement. The quantum mutual information is
generally taken to be the measure of total correlations, classical
and quantum, in a quantum state. For two systems, $A$ and $R$, it
is defined as
 \be
I(A:R) = H(A) + H(R) -H(A,R),
 \ee
where $H(\cdot)$ stands for the von Neumann entropy,
$H(\rho)\equiv -\tr(\rho\log\rho).$ In our paper, all logarithms
are taken to base 2. For a classical probability distribution,
Bayes' rule leads to an equivalent definition of the mutual
information as $I(A:R) = H(R)-H(R|A),$ where the conditional
entropy $H(R|A)$ is an average of the Shannon entropies of $R,$
conditioned on the alternatives of $A.$ It captures the ignorance
in $R$ once the state of $A$ has been determined. For a quantum
system, this depends on the measurements that are made on $A.$ If
we restrict to projective measurements described by a complete set
of projectors $\{\Pi_i\},$ corresponding to the measurement
outcome $i,$ the state of $R$ after the measurement is given by
    \be
\rho_{R|i} =
\tr_A(\Pi_i\rho_{AR}\Pi_i)/p_i,\;\;\;p_i=\tr_{A,R}(\Pi_i\rho_{AR}\Pi_i).
    \ee
A quantum analogue of the conditional entropy can then be defined
as $\tilde{H}_{\{\Pi_i\}}(R|A)\equiv\sum_ip_iH(\rho_{R|i}),$ and
an alternative version of the quantum mutual information can now
be defined as
 \be
\mathcal{J}_{\{\Pi_i\}}(A:R) = H(R)-\tilde{H}_{\{\Pi_i\}}(R|A).
 \ee
The above quantity depends on the chosen set of measurements
$\{\Pi_i\}.$ To capture all the classical correlations present in
$\rho_{AR},$ we maximize $\mathcal{J}_{\{\Pi_i\}}(A:R)$ over all
$\{\Pi_i\},$ arriving at a measurement independent quantity
$\mathcal{J}(A:R) =
\max_{\{\Pi_i\}}(H(R)-\tilde{H}_{\{\Pi_i\}}(R|A)) \equiv
H(R)-\tilde{H}(R|A),$ where
$\tilde{H}(R|A)=\min_{\{\Pi_i\}}\tilde{H}_{\{\Pi_i\}}(R|A).$ The
quantum discord is finally defined as
 \ben
 \label{discexp}
\mathcal{D}(A:R) &=& I(A:R)-\mathcal{J}(A:R) \\
                 &=&
                 H(A)-H(A:R)+\min_{\{\Pi_i\}}\tilde{H}_{\{\Pi_i\}}(R|A).\nonumber
 \een

As a first step towards the calculation of quantum discord, we
begin by rewriting the state $\rho_{AR}$ in a more conducive form,
as
 \ben
 \label{rhoardiff}
 \rho_{AR}&=&\frac{1-t^2}{2}\Big(\proj{0}\otimes M_{00}+\proj{1}\otimes M_{11} \nonumber\\
             && + \ket{0}\bra{1}\otimes M_{01} + \ket{1}\bra{0}\otimes
             M_{10}\Big),
 \een
where
 \ben
 \label{pieces}
 M_{00} &=& \sum_{n=0}^{\infty}t^{2n}\proj{n},\nonumber \\
 M_{11} &=& (1-t^2)\sum_{n=0}^{\infty}(n+1)t^{2n}\proj{n+1}, \nonumber \\
 M_{01} &=& \sqrt{1-t^2}\sum_{n=0}^{\infty}\sqrt{n+1}t^{2n}\ket{n}\bra{n+1},\nonumber \\
 M_{10} &=& M^{\dg}_{01}.
 \een
This form of the state suggests a natural bipartite split across
which to calculate the quantum discord. We have, in effect, a
$2\times \infty$ dimensional system, and we will make our
measurement on the 2 dimensional subsystem, which in our case,
will be Alice's side. It is now easy to obtain the reduced state
of the measured subsystem as
    \be
    \label{rhoa}
\rho_A = \tr_R(\rho_{AR})= \frac{1}{2}\left(
\begin{array}{cc}
  1 & 0 \\
  0 & 1 \\
\end{array}
\right),
    \ee
whereby $H(A)=1.$ The spectrum of the complete state $\rho_{AR}$
is given by
 \be
 \label{fullspec}
\bm{\lambda}(\rho_{AR})=
\left\{\frac{1-t^2}{2}t^{2n}(1+(n+1)(1-t^2))\right\}_{n=0}^{\infty},
 \ee
whereby
 \ben
 H(A:R)&=&-\frac{1-t^2}{2}\sum_{n=0}^{\infty}t^{2n}(1+(n+1)(1-t^2)) \\
        && \times\log\left(\frac{1-t^2}{2}t^{2n}(1+(n+1)(1-t^2))\right). \nonumber
 \een

\begin{figure}
 \resizebox{8.5cm}{6.5cm}{\includegraphics{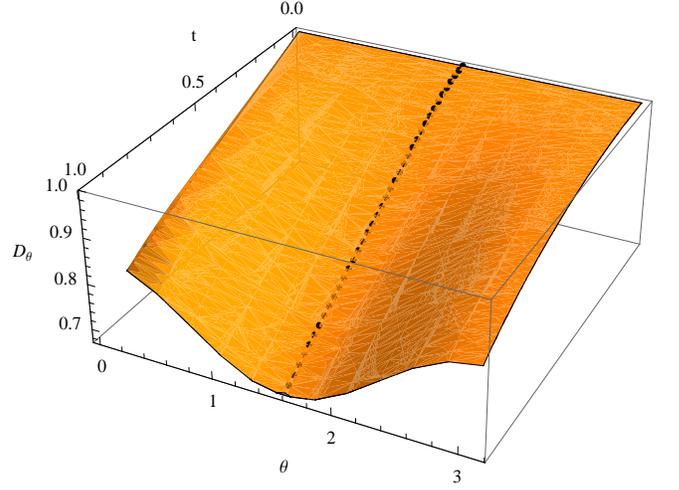}}
\caption{(Color online) The plot of the quantum discord,
$\mathcal{D}_{\theta},$~Eq. (\ref{dtheta}), as a function of
acceleration parameter $t=\tanh r$ and $\theta.$ In black dots are
shown the minima for different values of $t,$ which can be seen to
be attained for $\theta = \pi/2.$ They corresponds to the solid
green line in Fig. (\ref{discboson}).} \label{showmin}
\end{figure}

\begin{figure}
 \resizebox{7.5cm}{5cm}{\includegraphics{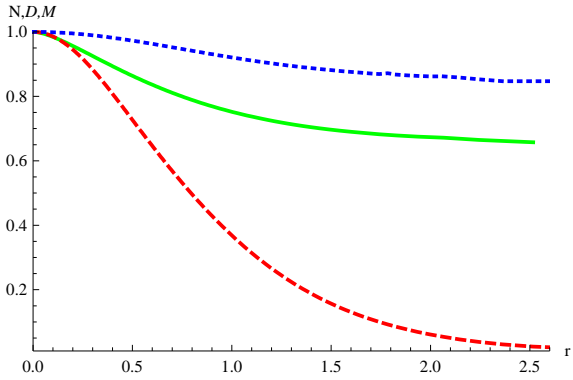}}
\caption{(Color online) The solid green line is the quantum
discord in the state $\rho_{AR}$ for a measurement made on Alice's
side. The red dashed line is the logarithmic negativity in the
same state, as in Ref.~\cite{sm05}. The blue dotted line is the
MID measure for the same state $\rho_{AR}$.} \label{discboson}
\end{figure}

        The evaluation of the quantum conditional entropy,
requires a minimization over all one-qubit projective
measurements, which are of the form
    \be
    \Pi_{\pm}=\frac{I_1\pm\bm{x}.\bm{\sigma}}{2}
    \ee
with $\bm{x}.\bm{x}=x_1^2+x_2^2+x_3^2=1,$ and $I_1$ is the
one-qubit, $2\times 2$ identity matrix. The post-measurement state
is then given by
 \ben
  \rho_{R|\pm}&=&\frac{1-t^2}{4p_{\pm}}\Big((1\pm x_3)M_{00} + (1\mp x_3)M_{11} \nonumber\\
                && \pm (x_1 + i x_2)M_{10}  \pm (x_1 - i x_2)M_{01}\Big),
 \een
with outcome probabilities
 $$
 p_{\pm} = \frac{1-t^2}{4}\left((1\pm x_3)\tr[M_{00}] + (1\mp x_3)\tr[M_{11}]\right) =
 \frac{1}{2}.
 $$
The density matrices $\rho_{R|\pm}$ are tridiagonal, whose
eigenvalues can be obtained easily numerically, in particular, by
using the parameterization $x_1=\sin\theta\cos\phi,$
$x_2=\sin\theta\sin\phi,$ and $x_3=\cos\theta.$ It is immediately
realized that the eigenvalues of these states, that are used to
calculate the conditional quantum entropy, are independent of
$\phi$. This is because the initial state in Eq. (\ref{rhoardiff})
is azimuthally invariant, and the final state whose spectrum is to
be evaluated reduces to
 \ben
  \rho_{R|\pm}&=&\frac{1-t^2}{2}\Big((1\pm \cos\theta)M_{00} + (1\mp \cos\theta)M_{11} \nonumber\\
                && \pm \sin\theta M_{10} \pm \sin\theta
                M_{01}\Big),
 \een
having spectra $\bm{\lambda}_{\pm}$. Then, following Eq.
(\ref{discexp}), the expression for quantum discord in the state
$\rho_{AR}$, as a function of the parameter $\theta,$ is given by
 \ben
 \label{dtheta}
 \mathcal{D}_{\theta}&=&1+\frac{1-t^2}{2}\sum_{n=0}^{\infty}t^{2n}(1+(n+1)(1-t^2)) \nonumber\\
        && \times\log\left(\frac{1-t^2}{2}t^{2n}(1+(n+1)(1-t^2))\right) \nonumber\\
        &-& \frac{1}{2}\sum_{i=\pm}\tr(\bm{\lambda}_i\log\bm{\lambda}_i),
 \een
and is plotted in Fig. (\ref{showmin}) as a function of $\theta$
and $t$. Realizing that the minimum is obtained for $\theta =
\pi/2,$ we obtained the final value of the quantum discord for the
state $\rho_{AR}$ as $\mathcal{D}=\mathcal{D}_{\theta=\pi/2}.$
This value is plotted in Fig. (\ref{discboson}). This is the main
result of our paper. To put our result in perspective, we also
plot the logarithmic negativity of the same state~\cite{sm05}.
This shows that, in the range of accelerations where the state has
no distillable entanglement, as shown by the vanishing logarithmic
negativity, the state indeed has finite quantum discord,

In the calculation of quantum discord, as per Eq.~(\ref{discexp}),
one maximizes over one-dimensional projective measurements on one
of the subsystems, in our case Alice. For the measurement-induced
disturbance (MID) measure, one performs measurements on
\emph{both} the subsystems, with the measurements being given by
projectors onto the eigenvectors of the reduced subsystems. This
can be thought of as a bidirectional form of discord, which
actually depends on the party making the measurement~\cite{wpm09}.
The MID measure of quantum correlations for a quantum state
$\rho_{AR}$ is given by~\cite{Luo08a}
 \be
 \mathcal{M}(\rho_{AR}) := \mathcal{I}(\rho_{AR}) - \mathcal{I}(\mathcal{P}(\rho_{AR}))
 \ee
where
 \be
 \label{E:midstate}
 \mathcal{P}(\rho_{AR}):=\sum_{i=1}^M\sum_{j=1}^N(\Pi_i^A\otimes\Pi_j^R)
\rho_{AR}(\Pi_i^A\otimes\Pi_j^R).
 \ee
Here $\{\Pi_i^A\},\{\Pi_j^R\}$ denote rank one projections onto
the eigenbases of $\rho_A$ and $\rho_R$, respectively.
$\mathcal{I}(\sigma)$ is the quantum mutual information, which is
considered to the measure of total, classical and quantum,
correlations in the quantum state $\sigma$. Since no optimizations
are involved in this measure, it is much easier to calculate in
practice than the quantum discord. The measurement induced by the
spectral resolution leaves the entropy of the reduced states
invariant and is, in a certain sense, the least disturbing.
Actually, this choice of measurement even leaves the reduced
states invariant~\cite{Luo08a}. For pure states, both the quantum
discord and the MID measure reduce to the von-Neumann entropy of
the reduced density matrix, which is a measure of bipartite
entanglement.

        Starting from the expression of $\rho_{AR}$ in Eq.
(\ref{rhoardiff}), we have
 \be
 \rho_{R}=\tr_{A}(\rho_{AR}) = \frac{1-t^2}{2}(M_{00}+M_{11}),
 \ee
which, being diagonal, leads to
 $$
 \{\Pi_j^R\}=\{E_j\}\;\;\;\;
 \mbox{where}\;\;\;[E_j]_{kl}=\delta_{kj}\delta_{lj},\;\;j,k,l=1,\cdots,\infty.
 $$
From Eq. (\ref{rhoa}),
 $$
 \{\Pi_j^A\}=\{E_j\}\;\;\;\;
 \mbox{where}\;\;\;[E_j]_{kl}=\delta_{kj}\delta_{lj},\;\;j,k,l=1,2.
  $$
Given these, $\mathcal{P}(\rho_{AR})=\mbox{diag}(\rho_{AR}) $ and,
 \ben
&& \hskip-0.7cm
 H(\mathcal{P}(\rho_{AR}))=-\frac{1-t^2}{2}\sum_{n=0}^{\infty}t^{2n}\log\left(\frac{1-t^2}{2}t^{2n}\right) -\frac{(1-t^2)^2}{2}\nonumber\\
           &&\hskip1.7cm \times\sum_{n=0}^{\infty}(n+1)t^{2n}\log\left((n+1)\frac{(1-t^2)^2}{2}t^{2n}\right)\nonumber\\
   &&\hskip-0.6cm =1 -
   \frac{3t^2}{1-t^2}\log(t)-\frac{3}{2}\log(1-t^2)-\frac{(1-t^2)^2}{2}\mathcal{S},
 \een
where $\mathcal{S}=\sum_{n=0}^{\infty}t^{2n}(n+1)\log(n+1).$ The
MID measure can now be calculated as
 \be
 \mathcal{M}(\rho_{AR}) = H(\mathcal{P}(\rho_{AR})) - H(\rho_{AR}),
 \ee
the latter of which can be obtained from Eq. (\ref{fullspec}). A
plot of this measure is shown in Fig. (\ref{discboson}), as the
blue dotted line.


We have shown the existence of purely quantum correlations between
two, initially entangled, free modes of a scalar field, when the
party detecting one of the modes undergoes a constant
acceleration, while the other is inertial. In this regime, there
there is no distillable entanglement between them as a consequence
of the Unruh effect. In particular, we provide evidence that there
is a finite amount of quantum discord in such a state in the limit
of infinite acceleration. As quantum discord captures nonclassical
correlations beyond entanglement, it might be possible to use
these correlations to attain nontrivial quantum advantage.


AD was supported in part by the EPSRC (Grant No. EP/C546237/1),
EPSRC QIP-IRC and the EU Integrated Project (QAP).

\vskip-0.5cm

\bibliography{rindler}

\end{document}